\documentclass[useAMS,usenatbib]{mn2e}
\usepackage{epsfig} 
\usepackage{graphicx}
\usepackage{amssymb}
\usepackage{natbib}
\usepackage{times}
\def\inte{{\em INTEGRAL}}
\def\XMM{{\em XMM-Newton}}

\def\chan{{\em Chandra}}
\def\swift{{\em Swift}}

\def\j16479{IGR\,J16479-4514}

\def\igrj11{IGR\,J11215-5952}

\def\2s{2S\,0114+65}
 
\def\j18483{IGR\,J18483-0311}
\def\saxj18{SAX\,J1818.6-1703}
\begin{document}

\title[IGR\,J18483-0311 in quiescence]{The supergiant fast X-ray transient IGR\,J18483-0311 in quiescence: \XMM,\ \swift,\ and \chan\ observations}

\author[A. Giunta, et al.]
{A. Giunta$^{1,2}$ \thanks{email: giunta@oa-roma.inaf.it}, 
E. Bozzo$^{3,1}$,   
F. Bernardini$^{1,2}$,   
G. Israel$^{1}$, 
L. Stella$^{1}$, 
M. Falanga$^{4}$, 
\newauthor 
S. Campana$^{5}$,
A. Bazzano$^{6}$,
A.J. Dean$^{7}$,  
M. Mendez$^{8}$\\
$^{1}$INAF - Osservatorio Astronomico di Roma, Via Frascati 33,
00044 Rome, Italy. \\
$^{2}$Dipartimento di Fisica - Universit\`a di Roma Tor Vergata, via
della Ricerca Scientifica 1, 00133 Rome, Italy.\\ 
$^{3}$ISDC, Geneva Observatory, University of Geneva, Chemin d'Ecogia 16, CH-1290 Versoix, Switzerland.\\
$^{4}$ISSI, Hallerstrasse 6, CH-3012 Bern, Switzerland.\\
$^{5}$INAF – Osservatorio Astronomico di Brera, via Emilio Bianchi 46, I-23807 Merate (LC), Italy.\\
$^{6}$INAF - IASF, Via del Fosso del Cavaliere 100, 00133 Roma, Italy.\\
$^{7}$School of Physics and Astronomy, University of Southampton, Southampton SO17 1BJ, UK.\\
$^{8}$Kapteyn Astronomical Institute, University of Groningen, PO Box 800, 9700 AV Groningen, the Netherlands. 
}

\date{Received 7 April 2009; accepted 28 May 2009.}

\maketitle

\begin{abstract} 
IGR\,J18483-0311 was discovered with \inte\ in 2003 and later classified as a supergiant fast X-ray transient. 
It was observed in outburst many times, but its quiescent state is still poorly known. 
Here we present the results of \XMM,\ \swift,\ and \chan\ observations of \j18483.\ 
These data improved the X-ray position of the source, and provided new information on the timing 
and spectral properties of  \j18483\ in quiescence. 
We report the detection of pulsations in the quiescent X-ray emission of this source, and give for the 
first time a measurement of the spin-period derivative of this source. 
In \j18483\ the measured spin-period derivative of -(1.3$\pm$0.3)$\times$10$^{-9}$~s~s$^{-1}$  
likely results from light travel time effects in the binary. 
We compare the most recent observational results of \j18483\ and \saxj18,\ the two supergiant 
fast X-ray transients for which a similar orbital period has been measured.    
\end{abstract}


\section{Introduction}
\label{sec:intro}

\j18483\ was discovered in 2003 during \inte\ deep observations of the Galactic 
Center \citep{chernyakova03}. The mean source X-ray flux was $\sim$10~mCrab  
in the 15-40 keV, and $\sim$5~mCrab in the 40-100 keV band \citep{chernyakova03, molkov04}. 
The 18.5~days orbital period of the system was discovered by \citet{levine06} using   
{\em RXTE} archival data, and was later confirmed with \inte\ \citep{sguera07}.    
\inte\ data also showed that \j18483\ usually undergoes relatively long outbursts  
($\sim$3~days) that comprise several fast flares with typical timescales of a few hours. 
During these bright events, the broad band (3-50~keV) spectrum is best fit by an 
absorbed cutoff power law model (photon index $\Gamma$=1.4, cut-off energy $E_{\rm c}$=22~keV, and 
absorption column density $N_{\rm H}$=9$\times$10$^{22}$~cm$^{-2}$). 
\citet{sguera07} further detected a periodicity at 21.0526$\pm$0.0005~s 
in the \inte\ data, and interpreted it as the spin-period of the neutron star (NS) hosted in 
\j18483.\ The measured pulse fraction in the 4-20~keV energy band was 48$\pm$7\%\footnote{Here we 
defined the pulsed fraction as $F$=($I_{\rm max}$-$I_{\rm min}$)/($I_{\rm max}$+$I_{\rm min}$), 
where $I_{\rm max}$ and $I_{\rm min}$ are the measured count rates at the maximum and at the minimum 
of the folded light curve, respectively.}. 
\swift\ /XRT observations in 2006 detected the source at a flux level of 4.2$\times$10$^{-12}$~erg/cm$^2$/s 
and provided a source position of  
$\alpha_{\rm J2000}$=18$^{\rm h}$48$^{\rm m}$17$\fs$17 and 
$\delta_{\rm J2000}$=-3${\degr}$10$\arcmin$15$\farcs$54 
\citep[estimated accuracy 3$\farcs$3,][]{sguera07}. This allowed \citet{rahoui08} to identify the 
optical counterpart of the X-ray source, a B0.5Ia star at a distance of 3-4~kpc, and to 
estimate its mass and radius ($M_{*}$=33~$M_{\odot}$ and 
$R_{*}$=33.8~$R_{\odot}$, respectively).  These authors also suggested 
that an eccentricity 0.43$\lesssim$$e$$\lesssim$0.68 could explain the 3-day duration of the outbursts 
\citep[as reported by][]{sguera07}. 

Based on these results, it was concluded that \j18483\ most likely belongs to the class of  
supergiant fast X-ray transients \citep[SFXTs,][]{sguera06, sguera07, walter07}. 
However, due to the longer duration of its outbursts (a few days as opposed to a 
few hours) and a factor of $\sim$10 lower luminosity swings between outburst and quiescence, 
\citet{rahoui08} classified \j18483\ as an ``intermediate'' SFXT, rather than a standard SFXT 
\citep[see][for the definition of standard and intermediate SFXTs]{walter07}.   

In this paper we analyze a 18~ks \XMM\ observation of \j18483\ in quiescence, and report the results of 
the spectral and timing analysis of this observation. 
We found that the pulse fraction of the source X-ray emission decreased significantly 
with respect to that measured while the source was in outburst, and provide for the first time 
an estimate of the spin-period derivative of this source.  
We also analyzed all the available \swift\ /XRT observations of \j18483,\ and studied  
the orbital variations of the source X-ray flux. 
A 1~ks \chan\ observation is also analyzed and provided an improved position of the X-ray source. 
The results from this study are then compared with those obtained recently on \saxj18,\ 
the other SFXT with a similar orbital period to that of \j18483.\     
So far, the orbital period has been measured with certainty only in other two SFXTs, i.e. 
IGR\,J16479-4514 \citep[3.3194~d][]{jain09}, and IGR\,J11215-5952 
\citep{romano07}\footnote{ However note that the behavior of IGR\,J11215-5952 is 
somewhat peculiar with respect to the other SFXTs, and thus \citet{walter07} 
excluded this system from their SFXT source list.}.   
\begin{figure}
\centering
\includegraphics[scale=0.34,angle=-90]{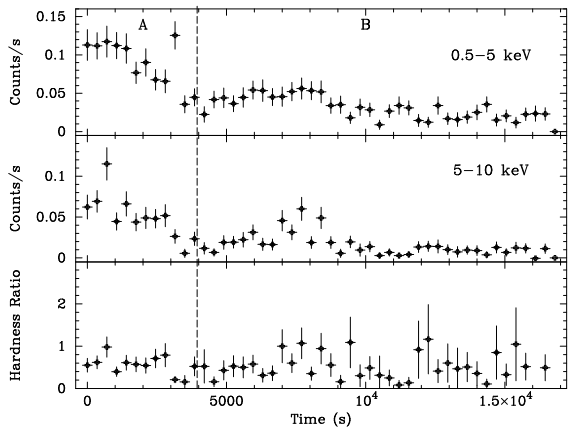}
\caption{ \XMM\ Epic-pn light curve of \j18483\ in the two energy bands 
0.5-5~keV and 5-10~keV (time bin is 350~s). The start time is 54020 (MJD) at 
9:22:04. The lower panel of the figure shows the hardness ratio (i.e. the ratio 
of the count-rate in the hard, 5-10~keV, and soft, 0.5-5~keV, energy band) versus time.}  
\label{fig:lcurve} 
\end{figure}

\section{ IGR\,J18483-0311: data analysis and results}
\label{sec:observation}
 
\subsection{ {\em XMM-Newton} data} 
\label{sec:xmm}
  
\XMM\ (\citealt{j01}) observed \j18483\ on 2006 October 12, 
and the total good exposure time was 14.4~ks 
(we discarded observational intervals  
that were affected by a high background). 
The observation data files (ODFs) were processed to produce calibrated
event lists using the standard \XMM\ Science Analysis
System (SAS 8.0). We used the {\sc epproc} and {\sc
emproc} tasks for the EPIC-PN and the two MOS
cameras, respectively. Source light curves and spectra were extracted in the
0.5--10~keV band, by using a circular extraction region with a radius of 20''. 
Background light curves and spectra were instead extracted by using a circular region 
with a radius of 50''.  
We used the SAS {\sc backscale} task and the {\sc lcmath} task in {\sc Heasoft} (version 6.6.1) 
to account for the difference in extraction areas between source and background spectra and 
light curves, respectively. 
The times of all light curves were corrected to the barycentre of the Solar System 
with the SAS {\sc barycen} task. 
In all cases, owing to poor statistics, the EPIC-MOS1 and EPIC-MOS2 cameras 
did not contribute significant additional information on the source spectra.  
Therefore, in the following we discuss only the spectra from the EPIC-PN camera. 
All spectra were rebinned in order to have at least 25 photons for each energy bin. 

\begin{figure}
\centering
\includegraphics[scale=0.34,angle=-90]{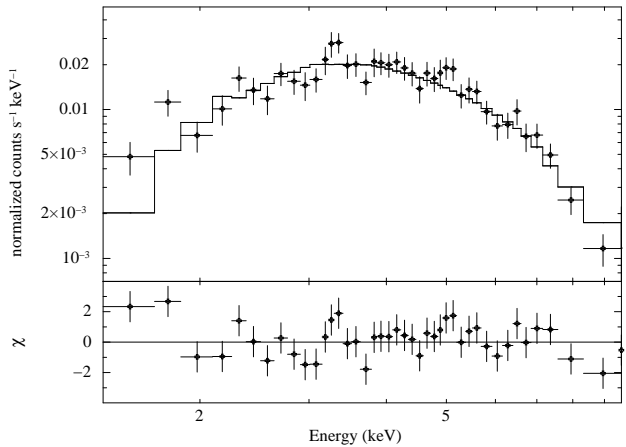}
\caption{ 0.5-10~keV \XMM\ Epic-pn spectrum of \j18483.\ The exposure time 
is 14.4~ks. The best fit is obtained using an absorbed power-law model (the best fit 
parameters are reported in the text). The lower panel of the figure 
shows the residuals from the best fit.}  
\label{fig:spectrum} 
\end{figure}

In Fig.~\ref{fig:lcurve} we report the X-ray light curves of the source in the 
0.5-5~keV and 5-10~keV energy bands; the lower panel of the figure shows 
the hardness ratio (i.e. the ratio of the count-rate in the hard, 5-10~keV, and soft, 0.5-5~keV, energy band) versus time.  
We note that the source count rate was decreasing during the first 5~ks of the 
observation. Unfortunately, the number of counts was insufficient to carry out any detailed investigation of the 
spectral variability. Therefore, we extracted only the 0.5-10~keV spectrum by 
using the total exposure time of the observation, and performed a fit with an absorbed power law model.  
The best fit parameters were $N_{\rm H}$=7.7$^{+1.2}_{-0.8}$$\times$10$^{22}$~cm$^{-2}$, and 
$\Gamma$=2.5$\pm$0.3 ($\chi^2_{\rm red}$/d.o.f.=1.3/39; hereafter 
errors are at 90\% confidence level, unless otherwise indicated). 
The absorbed (unabsorbed) flux in the 0.5-10~keV band was 9.3$\times$10$^{-13}$~erg/cm$^2$/s 
(5.2$\times$10$^{-12}$~erg/cm$^2$/s). Assuming a source distance of 4~kpc, the unabsorbed 
flux corresponds to an X-ray luminosity (0.5-10~keV) of 1.0$\times$10$^{34}$~erg/s.  
Figure~\ref{fig:spectrum} shows the Epic-PN source spectrum, 
together with the best fit model and the residuals from this fit. 
The 90\% confidence upper limit to the equivalent width for narrow iron lines at 6.4~keV and 
6.7~keV is 0.13~keV and 0.10~keV, respectively. 

\begin{figure}
\centering
\includegraphics[scale=0.35,angle=-90]{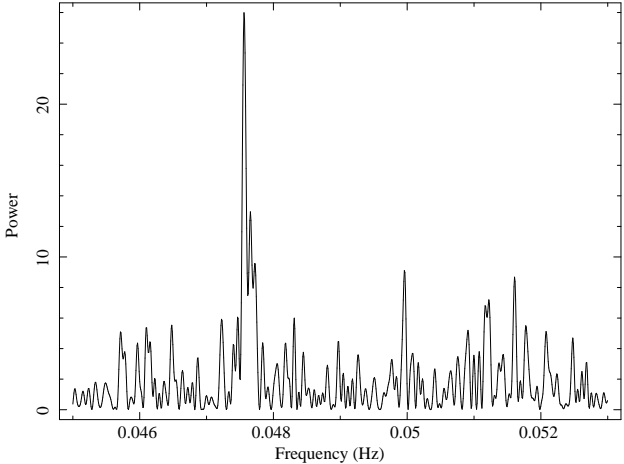}
\caption{The Z$^{2}_{1}$-statistic power spectrum of \j18483.\ We used the \XMM\ Epic-pn event file 
in the 0.5-10~keV energy band. The peak at Z$^{2}_{1}$$\sim$26 corresponds to a spin-period 
of 21.025$\pm$0.005~s}  
\label{fig:zstat} 
\end{figure}

Timing analysis of the \XMM\ data was carried out by using barycentred event files. 
We searched for the 21.0526~s spin-period of the NS in \j18483\ by performing  
first a power spectrum of the \XMM\ data. No significant evidence for a 
peak at the corresponding frequency was found. 
In order to investigate further the presence of pulsations in the \XMM\ data, 
we applied the Z$^{2}_{1}$-statistic technique \citep{buccheri88} to the photons event 
distribution for trial frequencies in a small window centered on 0.0475~Hz \citep{sguera07}. 
A spin-period of 21.025$\pm$0.005~s (hereafter errors on the NS spin 
periods are all at 1$\sigma$ confidence level) is found with a peak power of Z$^{2}_{1}$$\sim$26.  
The single-trial significance of this period is 4.7~$\sigma$. Figure~\ref{fig:zstat} shows 
the power spectrum computed with the Z$^{2}_{1}$-statistic technique 
by using the total exposure time of the \XMM\ observation. 
This period estimate was then refined by employing a phase fitting 
technique \citep[see e.g.,][]{osso03}. This gave our best determined spin period 
of $P_{\rm spin}$=21.033$\pm$0.004~s. In order to derive the significance of this result 
over the entire range of spin periods considered, we assumed a spin period derivative 
of 1.3$\times$10$^{-9}$~s/s (see Sect.~\ref{sec:discussion}) 
and multiplied the single trial significance of the Z$^{2}_{1}$-statistic for the total number 
of trial $D_{\rm P}$/($P_{\rm spin}^2$/2$Dt$). Here $D_{\rm P}$ is the separation in seconds 
between our measured spin period and that reported by \citet{sguera07}, and $Dt$=14.4~ks is 
the total duration of the \XMM\ observation. This gave us a significance of 3.7$\sigma$.  

\begin{figure}
\centering
\includegraphics[scale=0.34,angle=-90]{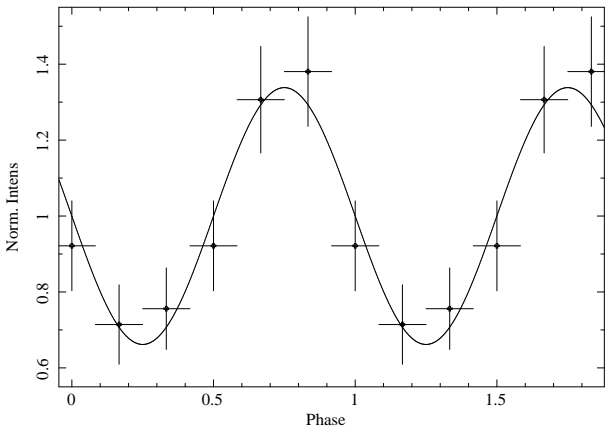}
\caption{Results of the epoch-folding analysis on the 
0.5-10~keV Epic-PN light curve of IGRJ18483-0311. We used the best determined 
source spin-period $P_{\rm spin}$=21.033$\pm$0.004~s.}  
\label{fig:phase} 
\end{figure}

From the folded light curve of the observation (obtained with the {\sc efold} task, see Fig.~\ref{fig:phase}), 
we measured a pulsed fraction of F=15$\pm$3\% in the 0.5-10~keV energy band. 
The profile is consistent with a sinusoid. 
In order to investigate the energy dependence of the pulse fraction, we also extracted 
and folded light curves in different energy bands and time intervals. We found that the pulsed fraction 
decreases slightly toward higher energies, whereas no significant variation could be measured 
across the ``A'' and ``B'' time intervals shown in Fig~\ref{fig:lcurve} (see Table~\ref{tab:pulse}). 
We note that the pulsed fraction we measured in the quiescent state of \j18483\ is a factor of $\sim$3   
lower than that reported by \citet{sguera07} during the source outburst\footnote{However, note that 
these pulse fractions are measured in slightly different energy bands (see Sect.~\ref{sec:intro}).}.   

\begin{table}
\caption{\j18483\ pulsed fractions (the time intervals ``A'' and ``B'' are 
indicated in Fig.~\ref{fig:lcurve}).}
\begin{tabular}{@{}llll@{}}
\noalign{\smallskip}
\hline
\hline
\noalign{\smallskip}
Energy Band & Total Obs. & A & B \\
\noalign{\smallskip}
\hline
\noalign{\smallskip}
0.5-10 keV & (15$\pm$3)$\%$ & $<$29$\%$ $^a$ & $<$32$\%$ $^a$ \\
0.5-5 keV & (23$\pm$3)$\%$ & (30$\pm$5)$\%$ & $<$33$\%$ $^a$ \\
5-10 keV & $<$18$\%$ $^a$ & $<$36$\%$ $^a$ & $<$29$\%$ $^a$ \\
\noalign{\smallskip}
\hline
\hline
\noalign{\smallskip}
\multicolumn{4}{l}{$^{a}$ 3$\sigma$ upper limit.}\\
\end{tabular}
\label{tab:pulse}
\end{table} 

\subsection{ {\em Swift} data} 
\label{sec:swift}

\begin{table*}
\caption{Log of the \swift\ observations of \j18483\ and \saxj18.\ Spectra are extracted for each observation 
in the Table, and fit with an absorbed power-law model (absorption column density N$_{\rm H}$ and photon index $\Gamma$). 
$F_{\rm unabs}$ is the XRT/PC unabsorbed flux in the 0.5-10~keV energy band. EXP 
indicates the total exposure time of each observation (~\swift\ observations comprise several snapshots 
and are not continuous pointings at the source).} 
\begin{tabular}{@{}lllllllll@{}}
\hline
\hline
\noalign{\smallskip}
 IGR\,J18483-0311\\
\hline
\noalign{\smallskip}
OBS ID & INSTR & START TIME & STOP TIME & EXP & $N_{\rm H}$ & $\Gamma$ & $F_{\rm unabs}$ & $\chi^2_{\rm red}$/d.o.f. \\
    &       &            &           &   ks  &  (10$^{22}$~cm$^{-2}$) &  & (erg/cm$^{2}$/s) \\
\noalign{\smallskip}
\hline
\noalign{\smallskip}
00035093001$^{\rm c}$ & XRT/PC & 2006-02-16 01:37:19 & 2006-02-16 22:36:57 & 7.9 & 6.0$^{+1.9}_{-1.6}$  & 1.4$^{+0.5}_{-0.4}$ & 5.3$\times$10$^{-11}$ & 1.3/21 \\

00035093002$^{\rm c}$ & XRT/PC & 2006-03-05 11:12:39 & 2006-03-05 17:58:56 & 5.6 & 5.0$\pm$1.0 & 1.2$\pm$0.3 & 6.4$\times$10$^{-11}$   & 1.1/51 \\

00035093003 & XRT/PC & 2008-09-26 13:49:39 & 2008-09-26 15:26:38 & 2.0 & 5.0 (fixed)         & 1.2 (fixed)        & 1.8$\times$10$^{-12}$   & --- \\
\noalign{\smallskip}
\hline
\noalign{\smallskip}
 SAX\,J1818.6-1703\\
\hline
\noalign{\smallskip}
00036128001 & XRT/PC & 2007-11-09 17:47:09 & 2007-11-09 21:04:57 & 1.6 & 6.0 (fixed) & 1.9 (fixed) & $<$2.1$\times$10$^{-12}$ $^{a,b}$ & --- \\

00036128003 & XRT/PC & 2008-04-18 14:38:56 & 2008-04-18 17:45:49 & 2.0 & 6.0 (fixed) & 1.9 (fixed) & 3.5$\times$10$^{-12}$ $^{a}$ & --- \\ 

00037889001 & XRT/PC & 2008-07-20 02:44:06 & 2008-07-21 01:21:56 & 3.7 &  6.0 (fixed) & 1.9 (fixed) & 8.5$\times$10$^{-12}$ $^{a}$ & --- \\ 
\noalign{\smallskip}
\hline
\hline
\multicolumn{9}{l}{$^{a}$ We determined the source count rate with {sosta} and used the spectral parameters given in \citet{zand06} within {\sc webpimms} to estimate }\\ 
\multicolumn{9}{l}{the source flux.} \\ 
\multicolumn{9}{l}{$^{b}$ 3$\sigma$ upper limit.} \\ 
\multicolumn{9}{l}{ $^{c}$ Previously reported by \citet{sguera07}.} \\ 
\end{tabular}
\label{tab:log} 
\end{table*} 

In Table~\ref{tab:log} we show a log of the \swift\ observations 
analyzed in the present study. Note that the observations ID 00035093001 and 
00035093002 were also published previously by \citet{sguera07}. 
We used the {\sc xrtpipeline} (v.0.12.1) 
task to process \swift\ /XRT data \citep[note that part of these data were  
published by][]{sguera08}. Standard event grades of 0-12
were selected for the XRT photon-counting (PC) mode;  
filtering and selection criteria were applied using 
{\sc ftools} ({\sc Heasoft} v.6.6.1). 
We created exposure maps through the {\sc xrtexpomap} task, and 
used the latest spectral redistribution 
matrices in the {\sc Heasarc} calibration database (v.011). 
Ancillary response files, accounting for different 
extraction regions, vignetting and PSF corrections, were generated 
using  the {\sc xrtmkarf} task.  
When required, we corrected PC data for pile-up, and used the 
{\sc xrtlccorr} to account for this correction in the 
background subtracted light curves. 
\begin{figure}
\centering
\includegraphics[scale=0.48]{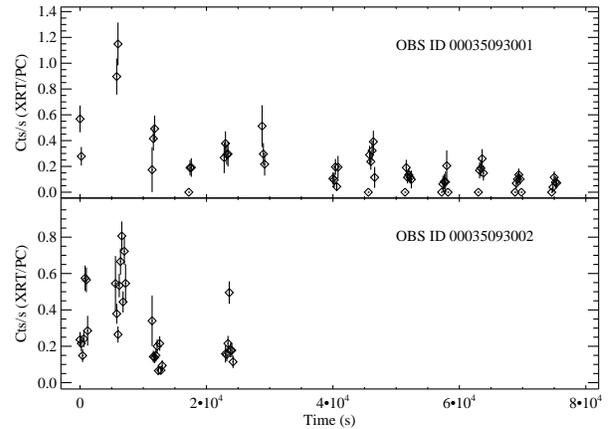}
\caption{\swift\ /XRT light curves of the observations ID.~00035093001 (upper panel), 
and ID.~00035093002 (lower panel). The start times of the light curves in the upper 
and lower panel are 53782.0695 (MJD) and 53799.4690 (MJD), respectively.}  
\label{fig:lcurve_swift}  
\end{figure}

For each observation in Table~\ref{tab:log}, we extracted the light curve and spectrum, 
and derived a mean X-ray flux by fitting this spectrum with an absorbed power-law model 
(we used {\sc Xspec} v.12.5.0). 
Spectra were rebinned in order to have at least 20 photons per bin and 
allow for $\chi^2$ fitting. 
In the observation ID.~00035093003 the very low source count rate 
(8.6$\pm$2.7$\times$10$^{-3}$) did not allow for a detailed 
spectral analysis. Therefore, we estimated the source count rate of the observation 
with {\sc sosta} ({\sc ximage} V.4.4.1), and then used this count rate within 
{\sc webpimms}\footnote{http://heasarc.nasa.gov/Tools/w3pimms.html}  
in order to derive the X-ray flux (we assumed the same spectral model 
of the observation ID.~00035093002).  

In Table~\ref{tab:log}, we report the best fit parameters obtained 
using an absorbed power law model to characterize the source spectra; 
Fig.\ref{fig:lcurve_swift} shows the source light curves from the observations
ID.~00035093001 and 00035093002 (note that, owing to poor statistics, we do not 
report the light curve of the observation ID.~00035093003). 
\begin{table}
\scriptsize
\caption{X-ray observations of IGR\,J18483-0311 and SAX\,J1818.6-1703 at 
different orbital phases.} 
\begin{tabular}{@{}lllll@{}}
\hline
\hline
\noalign{\smallskip} 
 IGR\,J18483-0311 & & & \\
\noalign{\smallskip}
\hline
\noalign{\smallskip}
INSTRUMENT & PHASE & $F_{\rm unabs}$$^a$ & $L_{\rm X}$$^c$ & ENERGY \\
           &       & (erg/cm$^{2}$/s) & (erg/s) & RANGE \\
\noalign{\smallskip}
\hline
\noalign{\smallskip}
IBIS/ISGRI$^1$ & 0 & 1.1$\times$10$^{-9}$ & 2.1$\times$10$^{36}$ & 20-100 \\
SWIFT/XRT$^{10}$ & 0.13 & 1.8$\times$10$^{-12}$ & 3.5$\times$10$^{33}$ & 0.5-10 \\
CHANDRA/HRC$^{10}$ & 0.24 & 4.3$\times$10$^{-10}$ & 8.3$\times$10$^{35}$ & 0.5-10 \\
XMM/Epic-pn$^{10}$ & 0.52 & 5.2$\times$10$^{-12}$ & 1.0$\times$10$^{34}$ & 0.5-10 \\
SWIFT/XRT$^{10}$ & 0.59 & 6.4$\times$10$^{-11}$ & 1.2$\times$10$^{35}$ & 0.5-10 \\
SWIFT/XRT$^{10}$ & 0.67 & 5.3$\times$10$^{-11}$ & 1.0$\times$10$^{35}$ & 0.5-10 \\
\noalign{\smallskip}
\hline
\hline
\noalign{\smallskip}
 SAX\,J1818.6-1703 & & & \\
\noalign{\smallskip}
\hline
\noalign{\smallskip}
INSTRUMENT & PHASE & $F_{\rm unabs}$$^a$ & $L_{\rm X}$$^d$   & ENERGY \\
           &       & (erg/cm$^{2}$/s) & (erg/s)  & RANGE \\
\noalign{\smallskip}
\hline
\noalign{\smallskip}
IBIS/ISGRI$^2$ & 0 & 3.8$\times$10$^{-10}$ & 2.9$\times$10$^{35}$ & 18-60 \\
IBIS/ISGRI$^4$ & 0.01 & 3.0$\times$10$^{-9}$ & 2.3$\times$10$^{36}$ & 18-60 \\
IBIS/ISGRI$^5$ & 0.06 & 3.8$\times$10$^{-10}$ & 2.9$\times$10$^{35}$ & 18-60 \\
SAX/WFC$^6$ & 0.07 & 2.1$\times$10$^{-9}$ & 1.6$\times$10$^{36}$ & 2-9 \\
SWIFT/XRT$^{10}$ & 0.12 & 3.5$\times$10$^{-12}$ & 2.6$\times$10$^{33}$ & 0.5-10 \\
SWIFT/XRT$^{10}$ & 0.22 & 8.5$\times$10$^{-12}$ & 6.4$\times$10$^{33}$ & 0.5-10 \\
XMM/Epic-pn$^7$ & 0.51 & $<$1.1$\times$10$^{-13}$$^b$ & $<$8.3$\times$10$^{31}$$^b$ & 0.5-10 \\
SWIFT/XRT$^{10}$ & 0.76 & $<$2.1$\times$10$^{-12}$$^b$ & $<$1.6$\times$10$^{33}$$^b$ & 0.5-10 \\
CHANDRA/ACIS-S$^{8,e}$ & 0.89 & 7.5$\times$10$^{-12}$ & 5.7$\times$10$^{33}$ & 0.5-10 \\
IBIS/ISGRI$^3$ & 0.91 & 3.0$\times$10$^{-10}$ & 2.3$\times$10$^{35}$ & 18-45 \\
IBIS/ISGRI+SWIFT/BAT$^2$ & 0.98 & $<$3.8$\times$10$^{-10}$$^b$ & $<$2.9$\times$10$^{35}$$^b$ & 18-60 \\
IBIS/ISGRI$^2$ & 1.00 & $<$3.8$\times$10$^{-10}$$^b$ & $<$2.9$\times$10$^{35}$$^b$ & 18-60 \\
IBIS/ISGRI+SWIFT/BAT$^2$ & 1.00 & 9.1$\times$10$^{-10}$ & 6.9$\times$10$^{35}$ & 18-60 \\
SWIFT/BAT$^9$ & 1.00 & 2.2$\times$10$^{-9}$ & 1.7$\times$10$^{36}$ & 15-150 \\
\noalign{\smallskip}
\hline
\hline
\noalign{\smallskip} 
\multicolumn{5}{l}{$^{a}$ throughout this paper we use the following conversion factors:} \\ 
\multicolumn{5}{l}{1 mCrab=2.08$\times$10$^{-11}$ erg/cm$^{2}$/s$^{1}$ (2-10 keV)} \\ 
\multicolumn{5}{l}{1 mCrab=7.57$\times$10$^{-12}$ erg/cm$^{2}$/s$^{1}$ (20-40 keV)} \\ 
\multicolumn{5}{l}{1 mCrab=9.42$\times$10$^{-12}$ erg/cm$^{2}$/s$^{1}$ (40-100 keV)} \\ 
\multicolumn{5}{l}{$^{b}$ 3$\sigma$ upper limit.} \\
\multicolumn{5}{l}{$^{c}$ From the unabsorbed flux and assuming a distance of 4~kpc (see Sect.~\ref{sec:intro}).} \\
\multicolumn{5}{l}{$^{d}$ From the unabsorbed flux and assuming a distance of 2.5~kpc \citep{masetti08}.} \\
\multicolumn{5}{l}{$^{e}$ Not corrected for absorption.} \\
\multicolumn{5}{l}{References: (1) \citet{sguera07}; (2) \citet{bird09};} \\
\multicolumn{5}{l}{(3) \citet{grebenev08}; (4) \citet{grebenev05};} \\
\multicolumn{5}{l}{(5) \citet{sguera05}; (6) \citet{zand98}; } \\
\multicolumn{5}{l}{(7) \citet{bozzo08c}; (8) \citet{zand06};} \\
\multicolumn{5}{l}{(9) \citet[][but see also http://gcn.gsfc.nasa.gov/notices\_s/306379/BA/]{barthelmy08};} \\ 
\multicolumn{5}{l}{(10) this work.} \\
\end{tabular}
\label{tab:comparison}
\end{table}

\subsection{ {\em Chandra} data}
\label{sec:chandra} 

\chan\ observed \j18483\ on 2008 February 19 for a total exposure 
time of 1.2~ks with the High Resolution 
Camera. We reduced these data using the {\sc ciao} software 
(v 4.1.1) and the latest calibration file available. 
The best source position is provided by the {\sc wavdetect} task at 
$\alpha_{\rm J2000}$=18$^{\rm h}$48$^{\rm m}$17$\fs$2 and 
$\delta_{\rm J2000}$=-3${\degr}$10$\arcmin$16$\farcs$8 
(the position accuracy is 0$\farcs$8 at a 90\% confidence level). 
This is perfectly in agreement with the optical position reported by \citet{rahoui08}. 
We also derived the source count rate (0.51$\pm$0.02~cts/s) of the observation 
and then used this count rate with {\sc webpimms} 
in order to estimate the source X-ray flux. The results are given  
in Table~\ref{tab:comparison} (we assumed the same spectral model 
of the \XMM\ observation).

\section{ SAX\,J1818.6-1703: data analysis and results} 

\subsection{{\em Swift data}}
\label{sec:igrj1818}

 \saxj18\ is the only SFXT whit an orbital period comparable 
to that of \j18483\ \citep[30$\pm$0.1~days,][]{bird09,zurita09}.  
In Table~\ref{tab:comparison} we report 
all observations of this source we found in the literature, together with 
two recent \swift\ /XRT observations that have not yet been published (ID.~00036128001 and 
00037889001). We analyzed all these \swift\ /XRT observations with the procedures 
described in Sect.~\ref{sec:swift} and 
reported the results in Table~\ref{tab:log}. Following \citet{bird09}, we measured  
in Table~\ref{tab:comparison} the orbital phase of each observation from the epoch 
of the outburst occurred on 53671~MJD (phase 0), 
so as to permit a comparison with the orbital changes of the X-ray flux in \j18483.\ 
This comparison is carried out in Sect.~\ref{sec:discussion}.

\section{Discussion and conclusions}
\label{sec:discussion} 

In this paper we reported on all available quiescent observations of \j18483,\  
one of the two SFXTs for which the spin and orbital periods  
have been measured with certainty (the other is IGR\,J11215-5952 with 
$P_{\rm spin}$=186.78~s, but see Sect.~\ref{sec:intro}). 
We report the detection of pulsations in the 
quiescent X-ray emission of this source, and give for the first time a measurement 
of its spin-period derivative.   
To our knowledge, the spin-period has so far been detected unambiguously 
in two other SFXTs \citep[$P_{\rm spin}$=228~s, and 
4.7~s in IGR\,J16465-4507 and IGR\,J1841.0-0536, respectively;][]{bamba01,walter07}; 
however, the orbital period of these sources is not known. 
On the contrary, in the case of \saxj18\ the orbital period is known, 
but the spin-period remains to be discovered. 

Recently it has been suggested that a measurement of the NS spin and orbital 
periods can be the key to distinguish between different models  
proposed for SFXT sources \citep{bozzo08}. 
In fact, all these models involve a NS accreting from the intense wind of its supergiant 
companion, but several different mechanisms have been invoked in order to explain 
the very large luminosity swings observed during their transitions between outburst 
and quiescence \citep{zand05, walter07}.  
In particular, \citet{bozzo08} suggested that, if very slow spinning NSs ($P_{\rm spin}$$\gtrsim$1000~s) 
in relatively close orbits (few tens of days) are hosted in SFXTs, 
then a magnetic gating mechanism can be invoked in order to explain such 
luminosity swings. In this case the NS magnetic field 
would be in the ``magnetar'' range \citep[i.e. $\gtrsim$10$^{14}$-10$^{15}$~G;][]{duncan92}. 
On the contrary, faster spin-periods   
might indicate that the large luminosity swings of SFXTs are caused by 
a centrifugal rather than a magnetic gating \citep[a similar mechanism was suggested to explain 
the pronounced activity of Be X-ray pulsar transient systems;][]{stella86}. 
Alternatively, the observed variations in the X-ray luminosity of SFXTs 
might also be caused by drastic changes in the mass accretion rate onto the NS  
due to an extremely clumpy wind or to large scale structure 
in the immediate surroundings of the supergiant companion. 
In these models, the orbital periods may be as high as hundreds of days  
\citep[see in particular][]{sidoli07,negueruela08}. 

In 2008, an \XMM\ observation of IGR\,J16479-4514 revealed that 
also eclipse-like events can contribute to the luminosity swings 
observed in SFXTs \citep{bozzo08b}. Therefore, besides a measurement of the NS 
spin and orbital period, also an in-depth monitoring of the X-ray flux 
and spectral changes at different orbital phases is required in order 
to distinguish between different models or scenarios for SFXT sources.  

To this aim, we presented in Table~\ref{tab:comparison} an analysis of the orbital 
changes in the X-ray flux observed from \j18483\ and \saxj18,\ the only two SFXTs 
with a comparable orbital period. 	 
In the case of \j18483\ only few observations have been carried out in quiescence and thus 
the orbital monitoring of this source is far from being complete 
\citep[following][ we measured the source phases from the epoch of the brightest 
outburst observed with \inte\ at 53844.2~MJD]{sguera07}. The lowest flux state of this source 
was caught by \swift\ /XRT at phase 0.13, i.e. relatively close to the orbital phase where the 
highest X-ray activity of the source has been observed in several occasions. Unfortunately, 
the poor statistics of this \swift\ /XRT observation prevented an accurate spectral 
analysis, and thus we could not investigate the origin of this low flux state. 
In the other two \swift\ /XRT observations a spectral analysis could be carried out, 
but we did not detect any indication of a significant spectral variation.  
Only in the \XMM\ observation we measured a slight increase in the  
spectral power law index. 
This suggests that X-ray flux changes in \j18483\ might have occurred due to 
genuine variations in the mass accretion rate onto the NS, rather than eclipse-like events.   
Note that, the detection of pulsations in the \XMM\ data 
are also in agreement with the accretion scenario\footnote{Pulsations in quiescence  
were also reported for other two SFXT sources, i.e. IGR\,J16465-4507 \citep{walter06} and 
AX\,J1841.0-0535 \citep{sidoli08}.}. This suggests that SFXTs undergo low level accretion even 
when they are not in outburst \citep[see also][]{sidoli07}. 

At odds with the case of \j18483,\ Table~\ref{tab:comparison} shows that the different orbital phases of  
\saxj18\ have been fairly well monitored. 
Unfortunately, the X-ray spectrum of this source could be well characterized only during the outburst, 
whereas in quiescence only the \chan\ observation provided a measurement of the spectral parameters 
(see Table~\ref{tab:log}). In all the other observations only a rough estimate of the source flux could be obtained.  
Note that the source was not detected by \XMM\ at the orbital phase 0.52,  
and the 3$\sigma$ upper limit on the source X-ray flux was at least 
an order of magnitude lower than the fluxes measured in any other orbital phases.  
Since no spectral analysis could be carried out on \saxj18\ at this orbital phase, 
the origin of this low flux event could not be investigated further. 
In case future observations of \saxj18\ reveal that this source regularly undergoes X-ray 
eclipses at the orbital phase $\sim$0.5, this can help clarifying the issue of the extreme flux 
changes in this source. 

More observations of \j18483\ and \saxj18\ at different orbital phases 
with high sensitivity X-ray telescopes, like \XMM\ and \chan,\ are clearly required 
in order to understand unambiguously the origin of their outburst/quiescent activity. 
Being these two sources the only SFXTs with a comparable orbital period, they are very well 
suited to test different models proposed to explain the behavior of SFXTs. 
We are currently investigating the results of the application of the gated accretion model 
to \j18483\ and \saxj18\ (Bozzo et al., 2009, in preparation). 

In this paper, besides X-ray flux changes, we also measured  
a spin-period variation in \j18483.\ By using our best-determined spin-period, 
$P_{\rm spin}$=21.033$\pm$0.004~s, and that found previously by \citet{sguera07}, 
we obtained a spin-period derivative in \j18483\ of 
$\dot{P}_{\rm spin}$=-(1.3$\pm$0.3)$\times$10$^{-9}$~s~s$^{-1}$. 
This value is comparable with the spin-period derivative measured in the case of the SFXT 
AX\,J1841.0-0535 \citep[-1.5$\times$10$^{-10}$~s/s][]{sidoli08} and those 
induced by accretion torques in wind-fed binaries 
\citep[see e.g.,][]{bildsten97}. However, in the present case we believe that the spin-period 
derivative most likely results from light travel time effects in the binary. 
In fact, in a binary system with an orbital period of $\sim$18.5~days, these effects can contribute 
to an apparent spin-period derivative of the order of 
$\sim$$v_{\rm orb}$/$c$=8.6$\times$10$^{-4}$~s~s$^{-1}$, i.e. much larger 
than the spin-period derivative we reported above \citep[here $v_{\rm orb}$ is the orbital 
velocity and $c$ is the light velocity; we used the mass and radius of the supergiant 
companion measured by][]{rahoui08}. Unfortunately, since a detailed orbital solution 
for this source is not yet available, we do not know if accretion torques acting onto the 
NS in \j18483\ might also have contributed to the observed spin-period derivative. 
Note that, in principle, this can be used to study the interaction between the NS and the 
inflowing matter from the supergiant companion \citep[see e.g.][]{bozzo08}.   

An orbital monitoring of \j18483\ is required in order to understand 
the origin of the measured spin-period derivative.

\section*{Acknowledgments} 
We thank the anonymous referee for his/her many helpful 
comments. EB thank M. Capalbi, M. Perri, and K. Page for 
kind help with the \swift\ /XRT data analysis. 
This work was partially supported 
through ASI and MIUR grants.

\end{document}